\pdfoutput=1

\documentclass[11pt,twoside,a4paper,cmspaper,final,collab]{cms-tdr}
\def\svnVersion{72206:72510}\def\cmsCernNo{2011-120}\def\cmsCernDate{\today}\def\cmsMessage{Submitted to Physical Review Letters}

\begin{document}\cmsNoteHeader{BPH-11-002}

\hyphenation{had-ron-i-za-tion}
\hyphenation{cal-or-i-me-ter}
\hyphenation{de-vices}
\RCS$Revision: 72510 $
\RCS$HeadURL: svn+ssh://alverson@svn.cern.ch/reps/tdr2/papers/BPH-11-002/trunk/BPH-11-002.tex $
\RCS$Id: BPH-11-002.tex 72510 2011-07-28 21:31:36Z alverson $
\def\vdef #1{\expandafter\def\csname #1\endcsname}
\def\vuse #1{\csname #1\endcsname}
\def\vu   #1{\csname #1\endcsname}
\providecommand{\PBzs}{\ensuremath{\PB^0_\cmsSymbolFace{s}\xspace}}
\providecommand{\cPBds}{\ensuremath{\cmsSymbolFace{B_{d(s)}}\xspace}}
\newcommand{\bsdll}{\ensuremath{\cPBds\to\ell^+\ell^-}\xspace}
\newcommand{\bsdmm}{\ensuremath{\cPBds\to\Pgmp\Pgmm}\xspace}
\newcommand{\bsmm}{\ensuremath{\PBzs\to\Pgmp\Pgmm}\xspace}
\newcommand{\bdmm}{\ensuremath{\PBz\to\Pgmp\Pgmm}\xspace}
\newcommand{\cbf}{\ensuremath{{\cal B}}}
\newcommand{\Bp}{\PBp\xspace}
\newcommand{\jpsi}{\cPJgy\xspace}
\newcommand{\Bs}{\PBzs\xspace}
\newcommand{\mup}{\Pgmp\xspace}
\newcommand{\mun}{\Pgmm\xspace}
\newcommand{\bupsik}{\ensuremath{\PBpm \to \cPJgy \PKpm}\xspace}
\renewcommand{\bspsiphi}{\ensuremath{\PBz \to \cPJgy \phi}\xspace}
\newcommand{\KS}{\PKS\xspace}
\newcommand{\Kp}{\PKp\xspace}
\newcommand{\Km}{\PKm\xspace}
\newcommand{\pim}{\Pgpm\xspace}
\newcommand{\pip}{\Pgpp\xspace}
\renewcommand{\KS}{\PKzS}
\newlength\figwid
\ifthenelse{\boolean{cms@external}}{\setlength\figwid{0.23\textwidth}}{\setlength\figwid{0.45\textwidth}}

\vdef{default-11:m2pt:0}   {\ensuremath{{4.0 } } }
\vdef{default-11:pt:0}   {\ensuremath{{6.5 } } }
\vdef{default-11:m1pt:0}   {\ensuremath{{4.5 } } }
\vdef{default-11:m2pt:0}   {\ensuremath{{4.0 } } }
\vdef{default-11:iso1:0}   {\ensuremath{{0.75 } } }
\vdef{default-11:chi2dof:0}   {\ensuremath{{1.6 } } }
\vdef{default-11:alpha:0}   {\ensuremath{{0.050 } } }
\vdef{default-11:alpha:1}   {\ensuremath{{0.025 } } }
\vdef{default-11:fls3d:0}   {\ensuremath{{15.0 } } }
\vdef{default-11:fls3d:1}   {\ensuremath{{20.0 } } }
\vdef{default-11:TRACKPTLO:cutValue}  {\ensuremath{{ 0.5 } } }
\vdef{default-11:TRACKETAHI:cutValue}  {\ensuremath{{ 2.4 } } }

\vdef{default-11:N-OBS-BPLUS0:sys}   {\ensuremath{{652 } } }
\vdef{default-11:N-OBS-BPLUS1:val}   {\ensuremath{{4450 } } }
\vdef{default-11:N-OBS-BPLUS1:sys}   {\ensuremath{{222 } } }

\vdef{default-11:N-EXP-SIG-BSMM0:val}   {\ensuremath{{ 1.19 } } }
\vdef{default-11:N-EXP-SIG-BSMM0:err}   {\ensuremath{{ 0.05 } } }
\vdef{default-11:N-EXP2-SIG-BSMM0:val}   {\ensuremath{{ 0.91 } } }
\vdef{default-11:N-EXP2-SIG-BSMM0:err}   {\ensuremath{{ 0.16 } } }
\vdef{default-11:N-EXP-SIG-BDMM0:val}   {\ensuremath{{0.076 } } }
\vdef{default-11:N-EXP-SIG-BDMM0:err}   {\ensuremath{{0.003 } } }
\vdef{default-11:N-EXP2-SIG-BDMM0:val}   {\ensuremath{{0.075 } } }
\vdef{default-11:N-EXP2-SIG-BDMM0:err}   {\ensuremath{{0.013 } } }

\vdef{default-11:N-EXP-SIG-BSMM1:val}   {\ensuremath{{ 0.55 } } }
\vdef{default-11:N-EXP-SIG-BSMM1:err}   {\ensuremath{{ 0.03 } } }
\vdef{default-11:N-EXP2-SIG-BSMM1:val}   {\ensuremath{{ 0.52 } } }
\vdef{default-11:N-EXP2-SIG-BSMM1:err}   {\ensuremath{{ 0.09 } } }
\vdef{default-11:N-EXP-SIG-BDMM1:val}   {\ensuremath{{0.030 } } }
\vdef{default-11:N-EXP-SIG-BDMM1:err}   {\ensuremath{{0.002 } } }
\vdef{default-11:N-EXP2-SIG-BDMM1:val}   {\ensuremath{{0.039 } } }
\vdef{default-11:N-EXP2-SIG-BDMM1:err}   {\ensuremath{{0.007 } } }

\vdef{default-11:N-PEAK-BKG-BS1:val}   {\ensuremath{{ 0.04 } } }
\vdef{default-11:N-PEAK-BKG-BS1:err}   {\ensuremath{{ 0.01 } } }
\vdef{default-11:N-PEAK-BKG-BD1:val}   {\ensuremath{{ 0.16 } } }
\vdef{default-11:N-PEAK-BKG-BD1:err}   {\ensuremath{{ 0.04 } } }
\vdef{default-11:N-EXP-BSMM0:val}   {\ensuremath{{ 0.60 } } }
\vdef{default-11:N-EXP-BSMM0:err}   {\ensuremath{{ 0.35 } } }

\vdef{default-11:N-PEAK-BKG-BS0:val}   {\ensuremath{{ 0.07 } } }
\vdef{default-11:N-PEAK-BKG-BS0:err}   {\ensuremath{{ 0.02 } } }
\vdef{default-11:N-PEAK-BKG-BD0:val}   {\ensuremath{{ 0.25 } } }
\vdef{default-11:N-PEAK-BKG-BD0:err}   {\ensuremath{{ 0.06 } } }

\vdef{default-11:N-EXP-BDMM0:val}   {\ensuremath{{ 0.40 } } }
\vdef{default-11:N-EXP-BDMM0:err}   {\ensuremath{{ 0.23 } } }
\vdef{default-11:N-EXP-BSMM1:val}   {\ensuremath{{ 0.80 } } }
\vdef{default-11:N-EXP-BSMM1:err}   {\ensuremath{{ 0.40 } } }
\vdef{default-11:N-EXP-BDMM1:val}   {\ensuremath{{ 0.53 } } }
\vdef{default-11:N-EXP-BDMM1:err}   {\ensuremath{{ 0.27 } } }

\vdef{default-11:N-OBS-BSMM0:val}   {\ensuremath{{2 } } }
\vdef{default-11:N-OBS-BDMM0:val}   {\ensuremath{{0 } } }
\vdef{default-11:N-OBS-BSMM1:val}   {\ensuremath{{1 } } }
\vdef{default-11:N-OBS-BDMM1:val}   {\ensuremath{{1 } } }
\cmsNoteHeader{BPH-11-002} 
\title{\texorpdfstring{Search for $\PBzs \to \mu^+ \mu^-$ and $\PBz \to \mu^+ \mu^-$ decays in $\Pp\Pp$ collisions at $\sqrt{s} = 7\,\mathrm{TeV}$}{Search for B(s) and B to dimuon decays in pp collisions at 7 TeV}}

\date{\today}

\abstract{A search for the rare decays $\PBzs \to \mu^+ \mu^-$ and $\PBz \to
\mu^+ \mu^-$ is performed in $\Pp\Pp$ collisions at $\sqrt{s}=7\,\mathrm{TeV}$, with a data sample corresponding to
an integrated luminosity of $1.14\,\mathrm{fb}^{-1}$, collected by the
CMS experiment at the LHC.
In both cases, the number of events observed after all selection
requirements is consistent with expectations from background and
standard-model signal predictions.
The resulting upper
limits on the branching fractions are ${\cal B}(\PBzs \to \mu^+ \mu^-)
< 1.9\times10^{-8}$ and ${\cal B}(\PBz \to \mu^+ \mu^-) <
4.6\times10^{-9}$, at 95\% confidence level.  }

\hypersetup{%
pdfauthor={CMS Collaboration},%
pdftitle={Search for B(s) and B to dimuon decays in pp collisions at 7 TeV},%
pdfsubject={CMS},%
pdfkeywords={CMS, physics}}

\maketitle

\vdef{lumi}{1.14}

\newcommand{\base}{default-11}
\newcommand{\noA}{NoData}
\newcommand{\noB}{NoData2011}
\newcommand{\noMc}{NoMc}
\newcommand{\sample}{NoData}
\newcommand{\samplet}{NoData}
\newcommand{\channel}{A}

In the standard model (SM) of particle physics, flavor-changing neutral
current (FCNC) decays are forbidden at tree level and can only proceed
through higher-order loop diagrams.  The decays \bsdll\ (where
$\ell=e,\mu$), besides involving $\cPqb \to \cPqs(\cPqd)$ FCNC
transitions through Penguin and box diagrams, are helicity suppressed by factors of
$(m_{\ell}/m_{\PB})^2$, where $m_{\ell}$ and $m_\PB$ are the masses of the
lepton and $\PB$ meson, respectively. They also require an internal
quark annihilation within the $\PB$ meson that further reduces the decay
rate by $(f_{\PB}/m_{\PB})^2$, where $f_\PB$ is the decay constant of the $\PB$
meson.

The SM-predicted branching fractions, $\cbf(\bsmm)=(3.2\pm
0.2)\times10^{-9}$ and
$\cbf(\bdmm)=(1.0\pm0.1)\times10^{-10}$~\cite{Buras:2010wr}, are
significantly enhanced in several extensions of the SM, although in
some cases the decay rates are lowered~\cite{Ellis:2007kb}.  For
example, in the minimal supersymmetric extension of the SM, the rates
are strongly enhanced at large values of
$\tan\beta$~\cite{Choudhury:2005rz,Parry:2005fp}.  In specific models
involving leptoquarks~\cite{Davidson:2010uu} and in supersymmetric
models with non-universal Higgs masses~\cite{Ellis:2006jy}, the \bsmm\
and \bdmm\ branching fractions can be enhanced by different factors
and, therefore, both channels must be studied in parallel. Several
experiments have published upper limits at 95\% confidence level (CL)
on these decays: $\cbf(\bsmm) < 5.1\times10^{-8}$ by
D0~\cite{Abazov:2010fs}; $\cbf(\bsmm) < 5.8\times10^{-8}$ and
$\cbf(\bdmm) < 1.8\times10^{-8}$ by CDF~\cite{Aaltonen:2007kv};
$\cbf(\bsmm) < 5.6\times10^{-8}$ and $\cbf(\bdmm) < 1.5\times10^{-8}$
by LHCb~\cite{Aaij:2011rj}.  CDF recently reported a new limit of
$\cbf(\bdmm) < 6.0\times10^{-9}$ and an excess of \bsmm\ events,
corresponding to $\cbf(\bsmm) =
(1.8^{+1.1}_{-0.9})\times10^{-8}$~\cite{CDF:2011fi}.

In this Letter, a simultaneous search for the \bsmm\ and \bdmm\ decays
is presented, using a data sample of $\Pp\Pp$ collisions at $\sqrt{s}
=7$\,TeV, corresponding to an integrated luminosity of $(\vuse{lumi}
\pm 0.07) \fbinv$, collected in the first half of 2011 by the Compact
Muon Solenoid (CMS) experiment at the Large Hadron Collider (LHC).  An
event-counting experiment is performed in dimuon mass regions around
the \Bs\ and \Bz\ masses.  To avoid any possible bias, the signal
region was kept blind until after all selection criteria were
established.  The backgrounds are evaluated from the yields measured
in data mass sidebands and from Monte Carlo (MC) simulations for rare
hadronic two-body $\PB$ decays.  The MC event samples are generated with
{\sc Pythia~6.409}~\cite{Sjostrand:2006za}, the unstable particles are
decayed via {\sc EvtGen}~\cite{Lange:2001uf}, and the detector
response is simulated with {\sc Geant4}~\cite{Ivanchenko:2003xp}.
Events of the type \bupsik, $\jpsi\to\mup\mun$ are used as a
normalization sample to minimize uncertainties related to the
$\bbbar$ production cross section and to the integrated luminosity.
The signal and normalization efficiencies are determined with MC
simulation studies.  A control sample of reconstructed \bspsiphi,
$\jpsi\to\mup\mun$ events is used to validate the MC simulation
(such as the \Bs\ transverse momentum \pt\ spectrum)
and to evaluate potential effects resulting from differences in
fragmentation between \Bp\ and \Bs.  The analysis is not affected by
multiple $\Pp\Pp$ collisions in the same bunch crossing (pileup)
because the spatial vertex resolution is good enough to correctly
identify
the $\Pp\Pp$ vertex from which signal candidates
originate. In the present data set, an average of 5.5 primary vertices
are reconstructed per event.

A detailed description of the CMS experiment can be found in
Ref.~\cite{:2008zzk}.  The main subdetectors used in this analysis are
the silicon tracker, composed of pixel and strip layers immersed in a
3.8\,T axial magnetic field, and the muon stations, made of
gas-ionization detectors embedded in the steel return yoke, and
divided into a barrel section and two endcaps.  The muons are tracked
within the pseudorapidity region $|\eta| < 2.4$, where $\eta =
-\ln[\tan(\theta/2)]$ and $\theta$ is the polar angle with
respect to the counterclockwise beam direction.  A muon \pt\
resolution of about 1.5\% is obtained for muons in this analysis.  The
events are selected with a two-level trigger system.  The first level
only requires two muon candidates, without an explicit
\pt\ requirement, while the high-level trigger (HLT) uses additional
information from the silicon tracker.
The HLT selection for the signal data sample requires two muons
each with $\pt>2\gev$, dimuon $\pt>4\gev$, invariant mass within $4.8 <
m_{\mu\mu} < 6.0\gev$, and a three-dimensional (3D) distance of
closest approach to each other $d'_\mathrm{ca} < 5\mm$.

The normalization (\bupsik) and control (\bspsiphi) samples were
collected with HLT requirements gradually tightened as the LHC
luminosity increased. This time evolution does
not affect the analysis presented here, which uses selection criteria
significantly tighter than any trigger requirements.  More than 95\%
of the normalization and control sample events were collected by
requiring two muons each with $\pt > 3\gev$, dimuon $\pt > 6.9\gev$,
invariant mass within $2.9 < m_{\mu\mu} < 3.3\gev$, $d'_\mathrm{ca} <
5\mm$, and a larger than 0.5\% probability of the $\chi^2$ per degree
of freedom (dof) of the dimuon vertex fit. Two additional trigger
requirements, measured in the transverse plane, significantly reduce
the rate of prompt \jpsi\ candidates: the significance of the flight
distance $\ell_{xy}/\sigma(\ell_{xy})$ must be larger than 3, where
$\ell_{xy}$ is the distance between the primary and dimuon vertices
and $\sigma(\ell_{xy})$ is its uncertainty; and the pointing angle
$\alpha_{xy}$ between the $\PB$ candidate momentum and the vector from
the primary vertex to the dimuon vertex must fulfill $\cos\alpha_{xy}
> 0.9$.  The average trigger efficiency for events in the signal and
normalization samples is about 80\%, as determined from MC simulation.
The uncertainty on the ratio of trigger efficiencies between the
signal and normalization samples is estimated to be 2\% by comparing
these ratios in simulation studies and in data.

Muon candidates are required to be reconstructed by two different
algorithms, one matching silicon-tracker tracks
to segments in the muon stations, and the other performing global
fits using tracks in
both detector systems~\cite{misid:2010}. The uncertainty on the ratio of muon
identification efficiencies between the signal and normalization
samples is estimated to be 5\%.

\begin{figure}[htbp]
\begin{centering}
\includegraphics[width=\figwid]{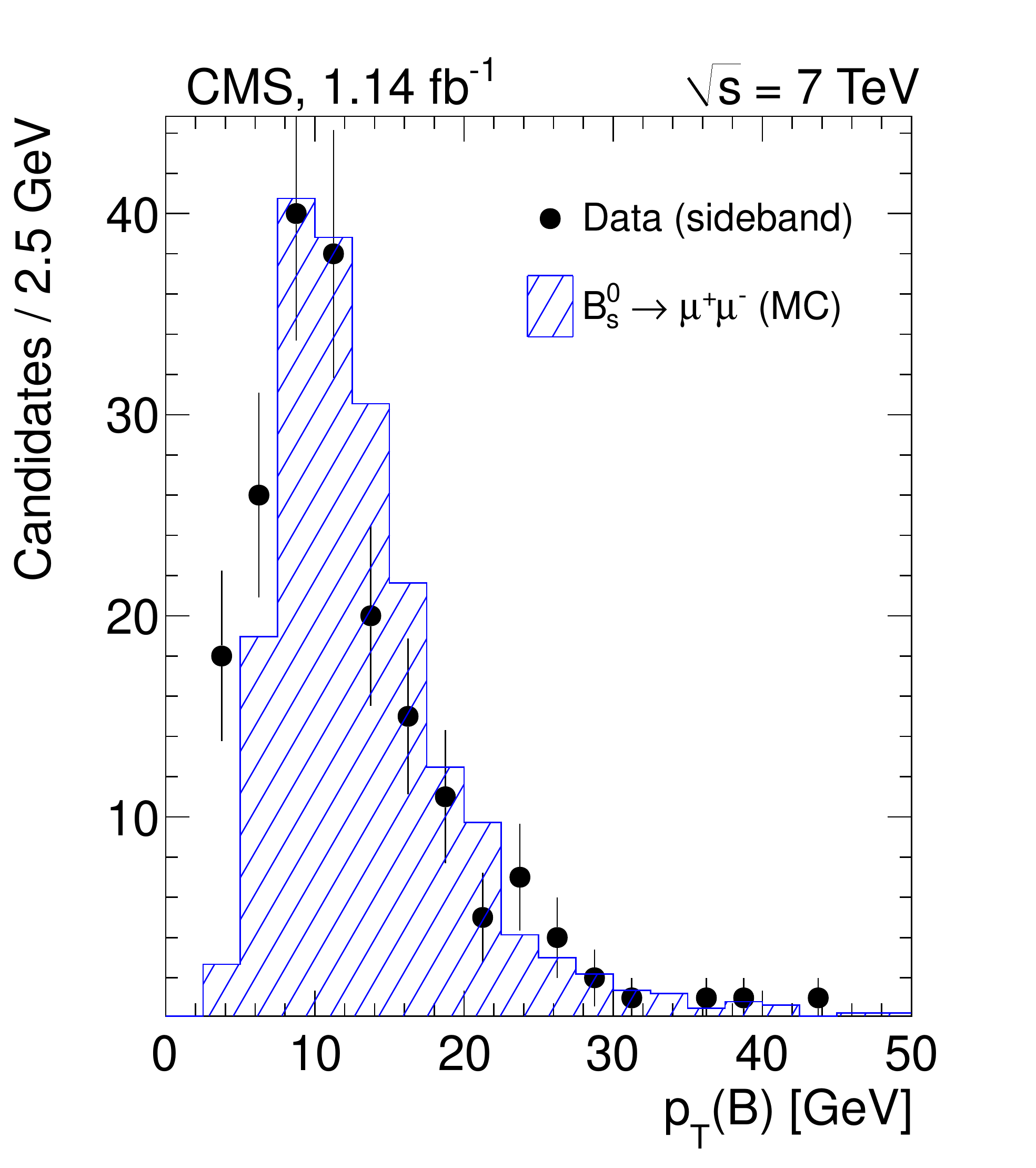}
\includegraphics[width=\figwid]{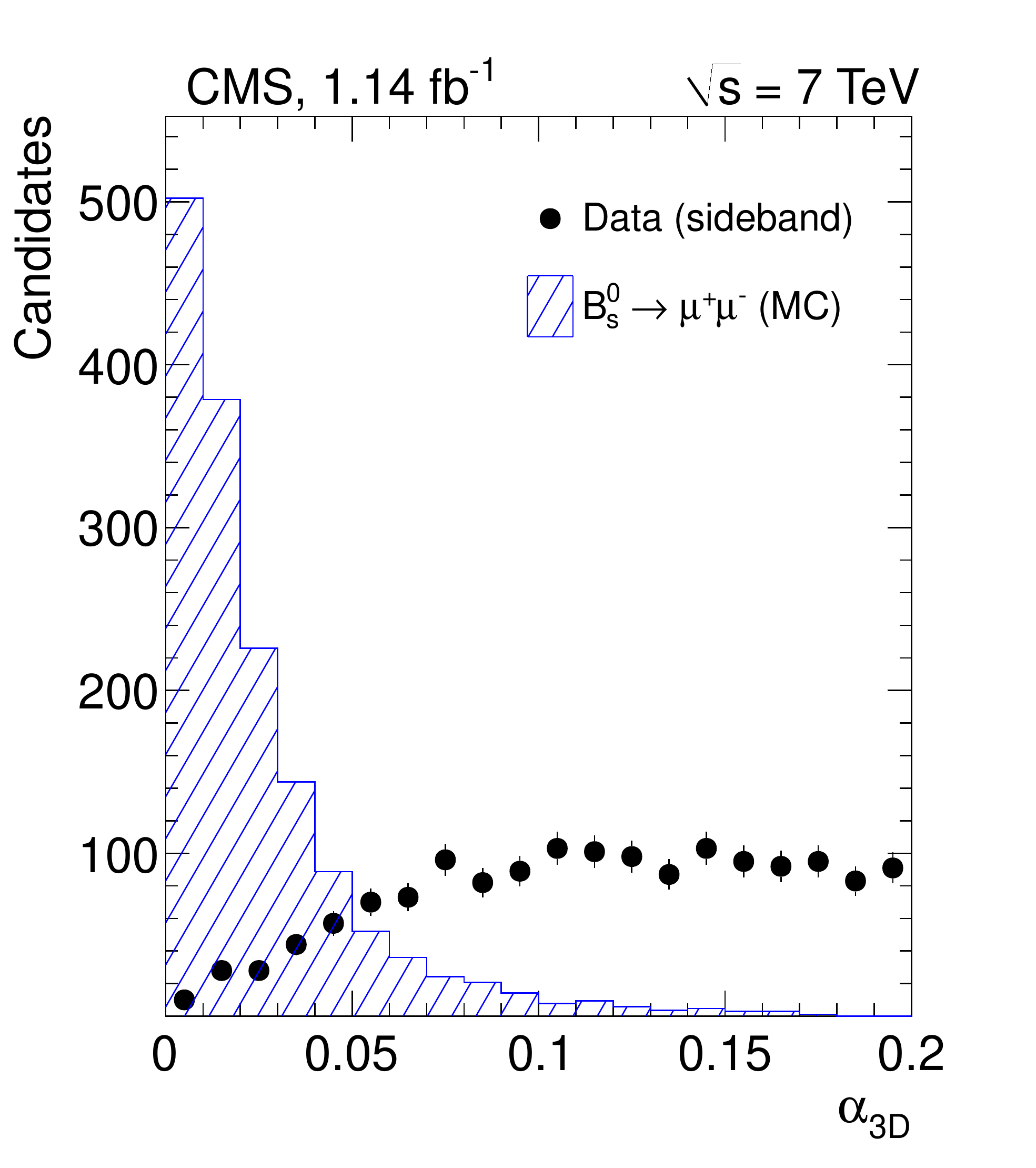}
\includegraphics[width=\figwid]{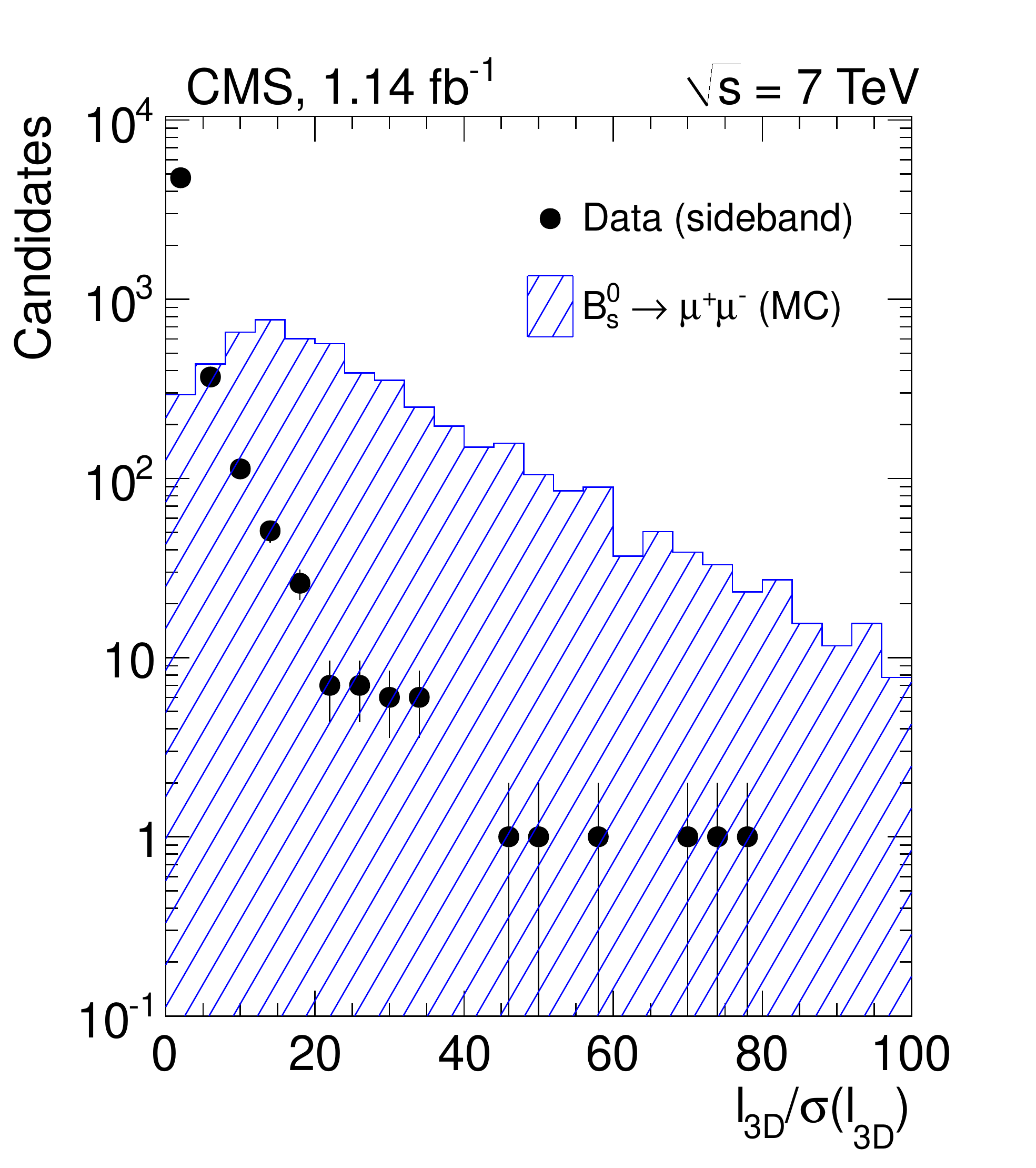}
\includegraphics[width=\figwid]{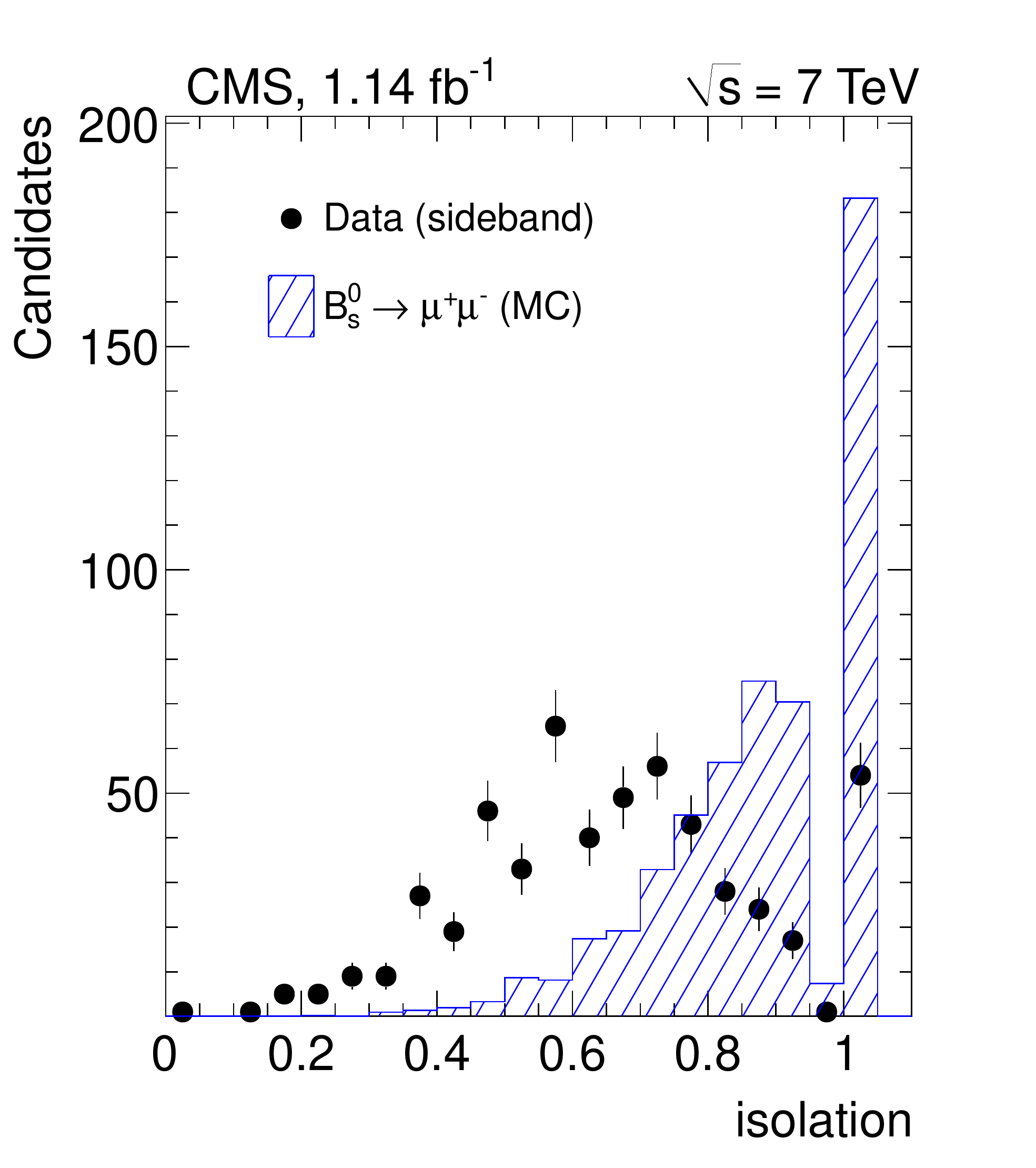}
    \caption{Comparison of MC signal and  sideband data distributions,
      for the transverse momentum (top left), the 3D pointing angle
      (top right), the flight length significance (bottom left), and
      the isolation (bottom right).  The MC histograms are
      normalized to the number of events in the data.}
    \label{fig:sgData-sgMc}
\end{centering}
\end{figure}

The $\PB\to\mup\mun$ candidates require two oppositely charged muons
with an invariant mass in the region $4.9 < m_{\mu\mu} < 5.9\gev$,
after constraining their tracks to come from a common vertex. The $\PB$
candidate momentum and vertex position are used to choose a primary
vertex based on the distance of closest approach.  Since the
background level depends significantly on the pseudorapidity of the
$\PB$ candidate, the events are separated into two categories: the
``barrel channel" contains the candidates where both muons have
$|\eta| < 1.4$ and the ``endcap channel" contains those where at least
one muon has $|\eta| > 1.4$.  An isolation variable $I =
\pt(\PB)/(\pt(\PB) + \sum_{\mathrm{trk}}\pt)$ is calculated from the
transverse momentum of the $\PB$ candidate $\pt(\PB)$ and the transverse
momenta of all other charged tracks satisfying $\Delta R =
\sqrt{(\Delta\eta)^2 + (\Delta\phi)^2} < 1$, where $\Delta\eta$ and
$\Delta\phi$ are the differences in pseudorapidity and azimuthal angle
between a charged track and the $\PB$ candidate momentum.  The sum
includes all tracks with $\pt>0.9\GeV$ that are consistent with
originating from the same primary vertex as the $\PB$ candidate or have
a distance of closest approach $d_\mathrm{ca} < 0.5\mm$ with respect
to the $\PB$ vertex. The minimum distance of closest approach with
respect to the $\PB$ vertex among all tracks in the event,
$d^{\mathrm{min}}_\mathrm{ca}$, is also determined as a complementary
isolation variable. Figure~\ref{fig:sgData-sgMc} illustrates the
transverse momentum, the 3D pointing angle $\alpha_{\mathrm{3D}}$, the 3D
flight length significance
$\ell_{3\mathrm{D}}/\sigma(\ell_{3\mathrm{D}})$, and the isolation
distributions for signal MC and for sideband background data events.
The sideband covers the range $4.9 < m_{\mu\mu} < 5.9\gev$, excluding
the signal window $5.2 < m_{\mu\mu} < 5.45\gev$.

The following selection requirements were optimized for the best
expected upper limit using MC signal events and data sideband
events. The requirements were established before observing the number
of data events in the signal region. The optimized requirements
include $\pt > \vuse{default-11:m1pt:0} \GeV$ on one muon and $\pt >
\vuse{default-11:m2pt:0} \gev$ on the other,
$\PB$ candidate $\pt > \vuse{default-11:pt:0} \gev$, $I >
\vuse{default-11:iso1:0} $, and $\PB$-vertex fit
$\chi^2/\mathrm{dof}<\vuse{default-11:chi2dof:0} $.  Two requirements
are different for the barrel and endcap channels: $\alpha_{3\mathrm{D}} <
\vuse{default-11:alpha:0} $ ($\vuse{default-11:alpha:1} $) and
$\ell_{3\mathrm{D}}/\sigma(\ell_{3\mathrm{d}}) >\vuse{default-11:fls3d:0} $
($\vuse{default-11:fls3d:1} $) for the barrel (endcap).
Furthermore, for events in the endcap
there is an additional
requirement, $d^{\mathrm{min}}_\mathrm{ca} > 0.15\mm$.
The signal efficiencies
$\varepsilon_{\mathrm{tot}}$ of these selections are provided in
Table~\ref{t:allTheNumbers}.  The dimuon mass resolution for signal
events depends on the pseudorapidity of the $\PB$ candidate and ranges
from $36\MeV$ for $\eta\approx0$, to $85\MeV$ for $|\eta| > 1.8$, as
determined from simulated signal.

\renewcommand{\base}{default-11}
\newcommand{\StrutSmall}{\rule[-1.0mm]{0pt}{4.7mm}}
\newcommand{\StrutLarge}{\rule[-2.3mm]{0pt}{6.3mm}}

\begin{table*}
  \begin{center}
    \caption{The event selection efficiencies for signal events
 $\varepsilon_{\mathrm{tot}}$, the SM-predicted number of signal events
 $N_{\mathrm{signal}}^{\mathrm{exp}}$, the expected number of
 combinatorial background events $N_{\mathrm{comb}}^{\mathrm{exp}}$
 and peaking background events $N_{\mathrm{peak}}^{\mathrm{exp}}$, and
 the number of observed events $N_{\mathrm{obs}}$ in the barrel and
 endcap channels for \bsmm\ and \bdmm.  }
    \label{t:allTheNumbers}
    \vspace{0.01in}
    \begin{tabular}{|l|c|c|c|c|c|c|}
      \hline
      &\multicolumn{2}{c|}{Barrel} &\multicolumn{2}{c|}{Endcap} \\
      \hline
      &\bdmm &\bsmm &\bdmm &\bsmm\StrutSmall
      \\
      \hline
      $\varepsilon_{\mathrm{tot}}$\StrutLarge
      &$(3.6\pm0.4)\times10^{-3}$
      &$(3.6\pm0.4)\times10^{-3}$
      &$(2.1\pm0.2)\times10^{-3}$
      &$(2.1\pm0.2)\times10^{-3}$
      \\
      $N_{\mathrm{signal}}^{\mathrm{exp}}$\StrutLarge
&$0.065\pm  0.011$
&$0.80 \pm  0.16$
&$0.025  \pm 0.004$
&$0.36 \pm  0.07$
      \\
      \hline
      $N_{\mathrm{comb}}^{\mathrm{exp}}$\StrutLarge
      &$\vuse{\base:N-EXP-BDMM0:val} \pm \vuse{\base:N-EXP-BDMM0:err}  $
      &$\vuse{\base:N-EXP-BSMM0:val} \pm \vuse{\base:N-EXP-BSMM0:err} $
      &$\vuse{\base:N-EXP-BDMM1:val} \pm \vuse{\base:N-EXP-BDMM1:err}  $
      &$\vuse{\base:N-EXP-BSMM1:val} \pm \vuse{\base:N-EXP-BSMM1:err}  $
      \\
      $N_{\mathrm{peak}}^{\mathrm{exp}}$\StrutLarge
      &$\vuse{\base:N-PEAK-BKG-BD0:val} \pm \vuse{\base:N-PEAK-BKG-BD0:err} $
      &$\vuse{\base:N-PEAK-BKG-BS0:val} \pm \vuse{\base:N-PEAK-BKG-BS0:err} $
      &$\vuse{\base:N-PEAK-BKG-BD1:val} \pm \vuse{\base:N-PEAK-BKG-BD1:err} $
      &$\vuse{\base:N-PEAK-BKG-BS1:val} \pm \vuse{\base:N-PEAK-BKG-BS1:err} $
      \\
      \hline
      \hline
      $N_{\mathrm{obs}}$
      &$\vuse{\base:N-OBS-BDMM0:val} $
      &$\vuse{\base:N-OBS-BSMM0:val} $
      &$\vuse{\base:N-OBS-BDMM1:val} $
      &$\vuse{\base:N-OBS-BSMM1:val} $
      \\
      \hline
    \end{tabular}
  \end{center}
\end{table*}

\renewcommand{\sample}{SgData-SgMc}
\renewcommand{\channel}{A}

\begin{figure}[htbp]
  \begin{centering}
\includegraphics[width=\figwid]{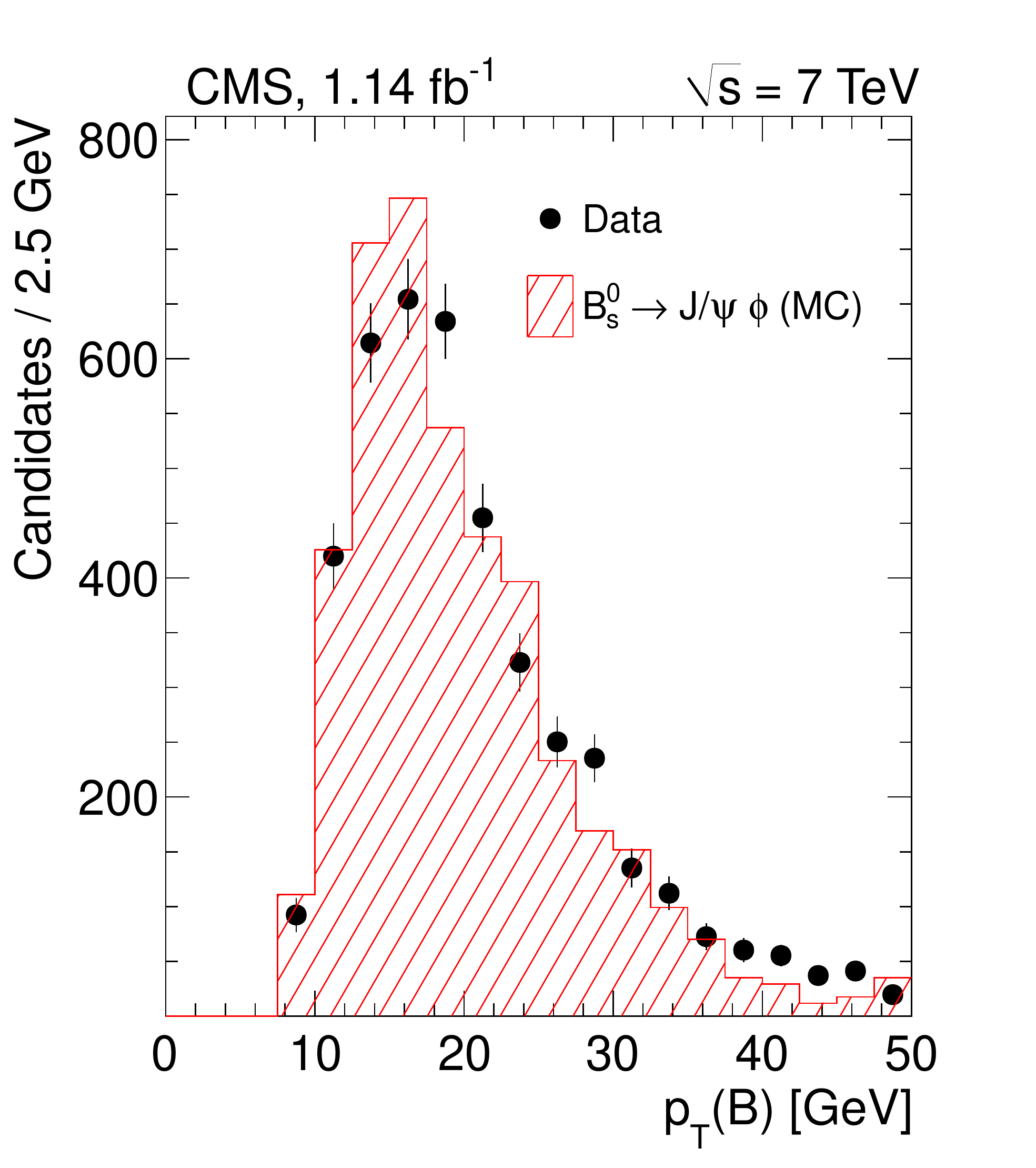}
\includegraphics[width=\figwid]{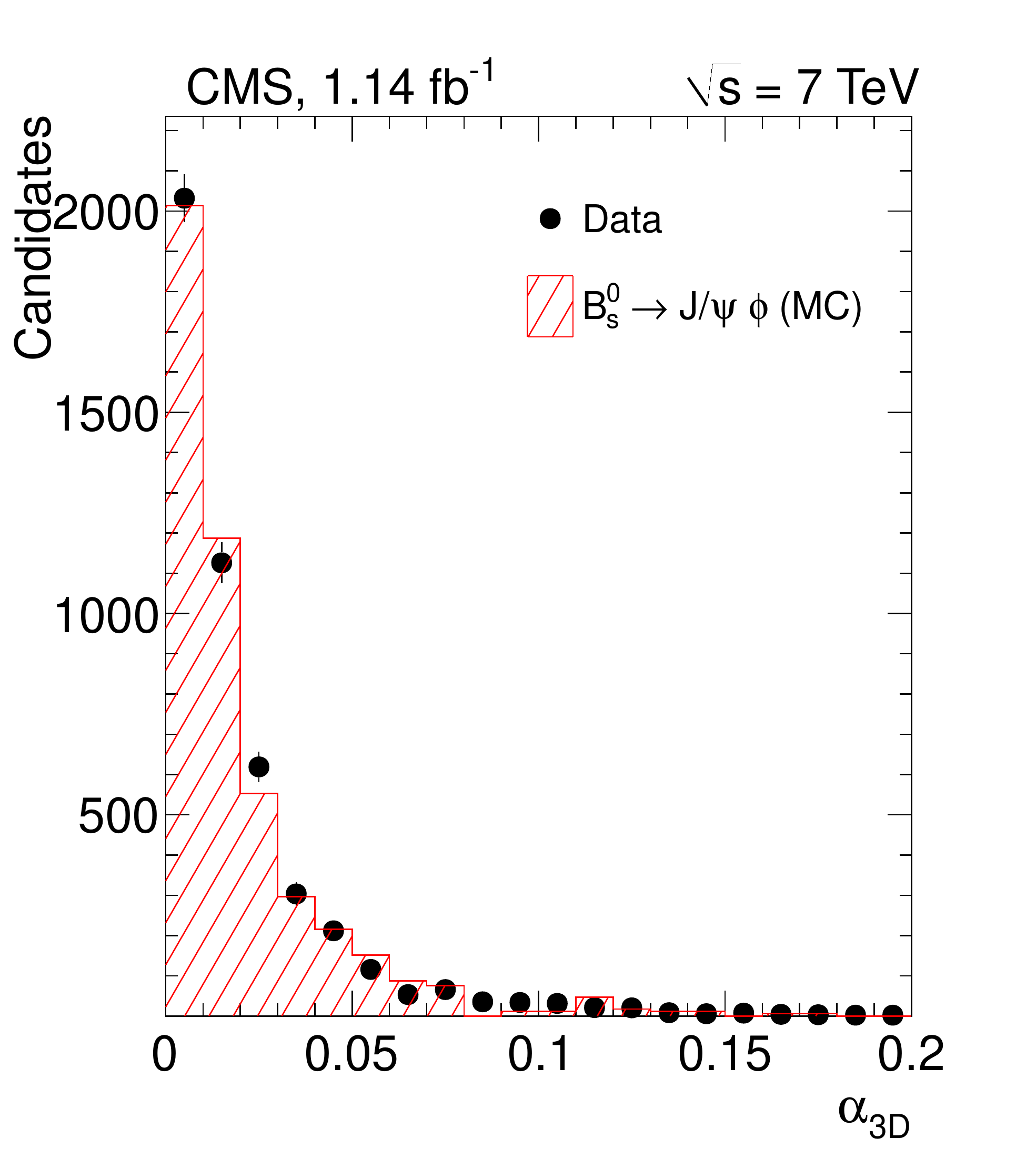}
\includegraphics[width=\figwid]{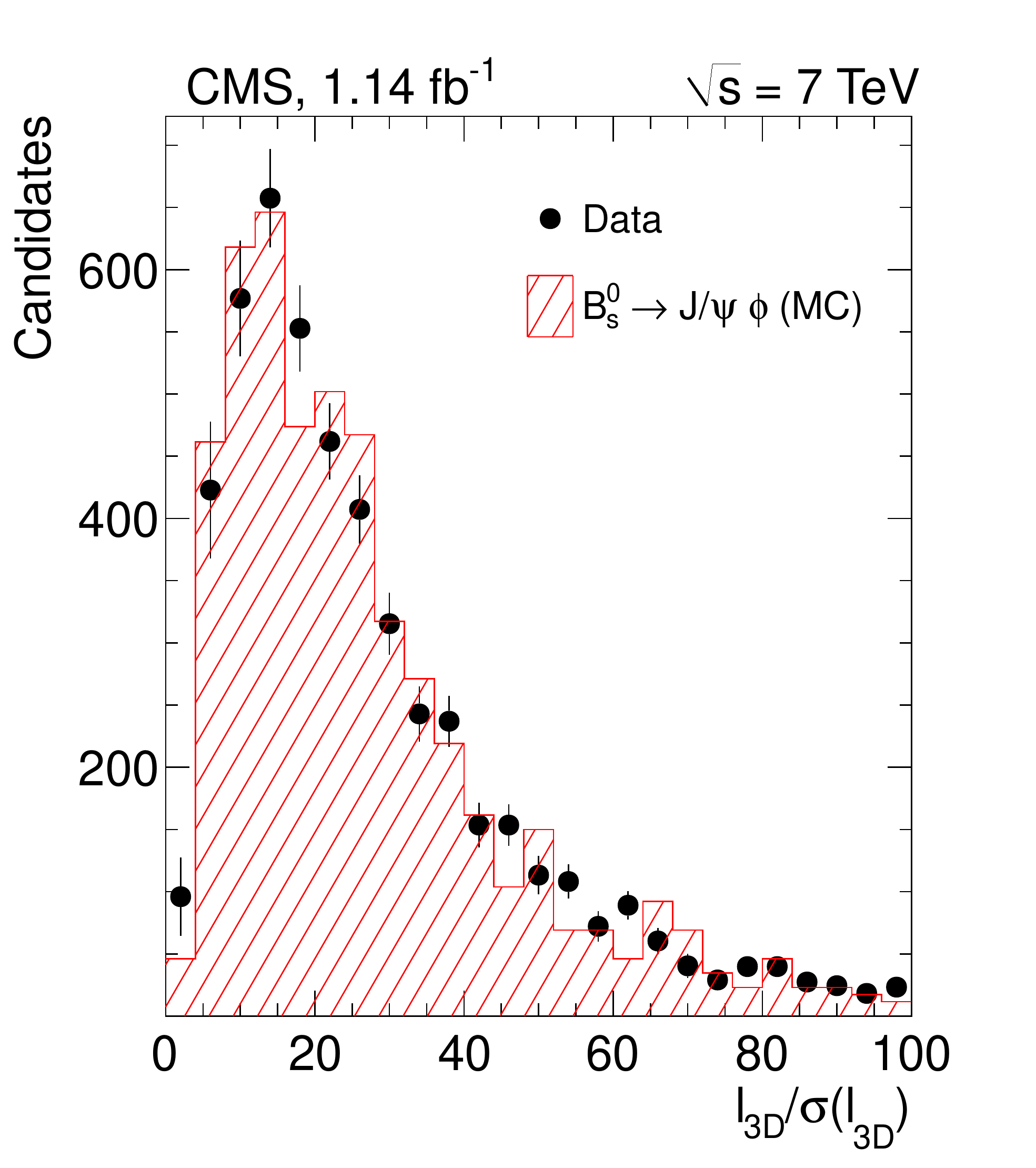}
\includegraphics[width=\figwid]{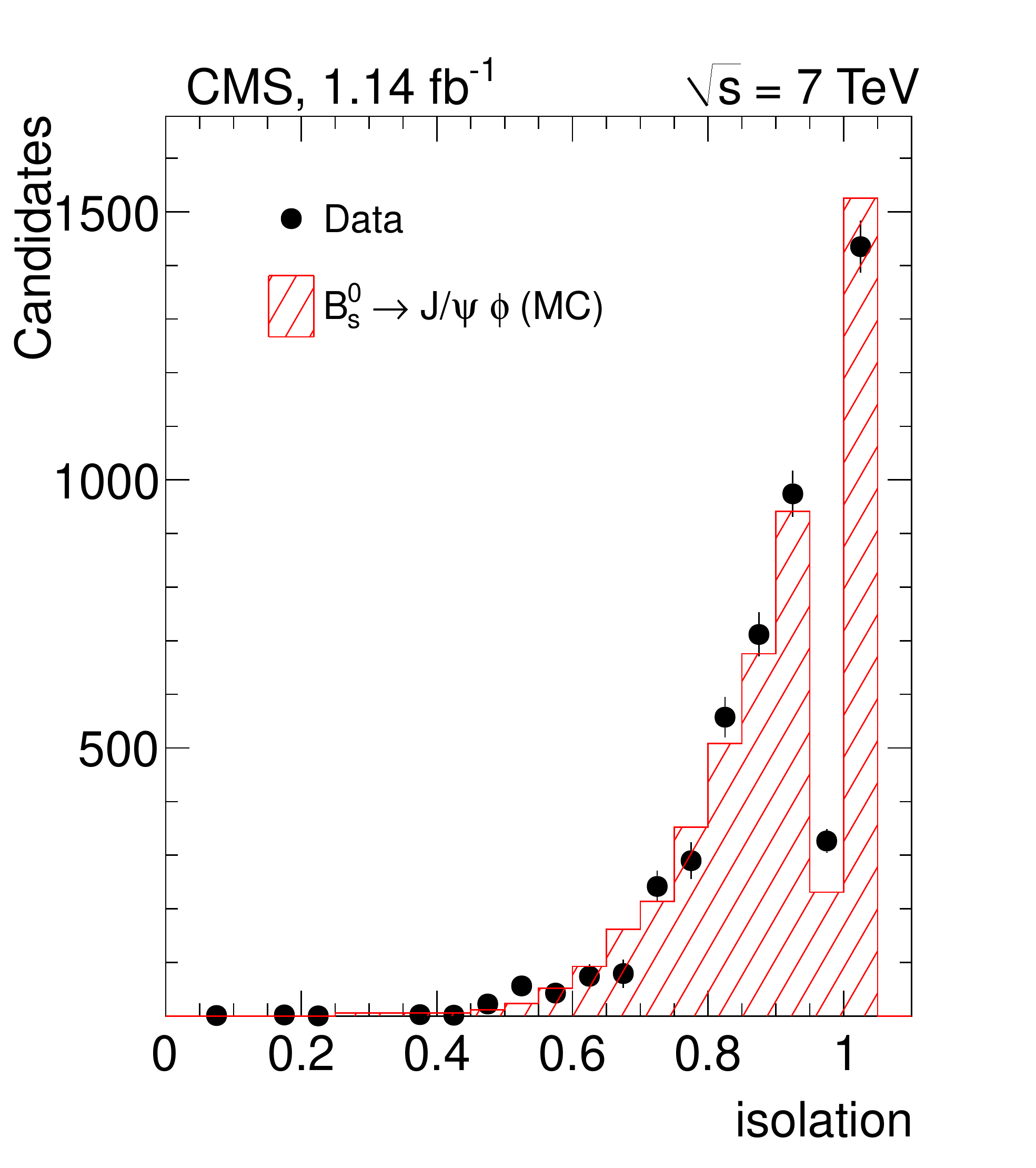}
    \caption{Comparison of measured and simulated \bspsiphi\
     distributions, for the transverse momentum (top left), the 3D
      pointing angle (top right), the flight length significance
      (bottom left), and the isolation (bottom right).  The MC
      histograms are normalized to the number of events in the data.}
   \label{fig:CsData-CsMc}
  \end{centering}
\end{figure}

The reconstruction of $\bupsik\to\mup\mun \PKpm$
($\bspsiphi\to\mup\mun\PKp\PKm$) candidates requires two
oppositely-charged muons with an invariant mass in the range
3.0--3.2\gev, which are combined with one (two) track(s), assumed to
be (a) kaon(s), fulfilling $\pt > \vuse{default-11:TRACKPTLO:cutValue}
\gev$ and $|\eta| < \vuse{default-11:TRACKETAHI:cutValue} $.  To
ensure a well-measured trigger efficiency, the selected candidates
must have dimuon $\pt > 7\gev$ and the two muons must bend away from
each other in the magnetic field (to avoid spurious detector-induced
pair correlations).  The $d'_\mathrm{ca}$ between all pairs among the
three (four) tracks is required to be less than $1\mm$.  For
\bspsiphi\ candidates the two assumed kaon tracks must have an
invariant mass in the range 0.995--1.045\gev and $\Delta R(\Kp,\Km) <
0.25$.  The tracks from all decay products are used in the $\PB$-vertex
fit and only $\PB$ candidates with an invariant mass in the range
4.8--6.0\gev are considered.  The efficiencies of individual selection
criteria agree to better than 4\% (6\%) between data and MC simulation
for the normalization (control) sample.  Figure~\ref{fig:CsData-CsMc}
compares
several distributions for \bspsiphi\ candidates between MC
simulation and sideband-subtracted data.

\begin{figure}[htbp]
  \begin{centering}
\includegraphics[width=\figwid]{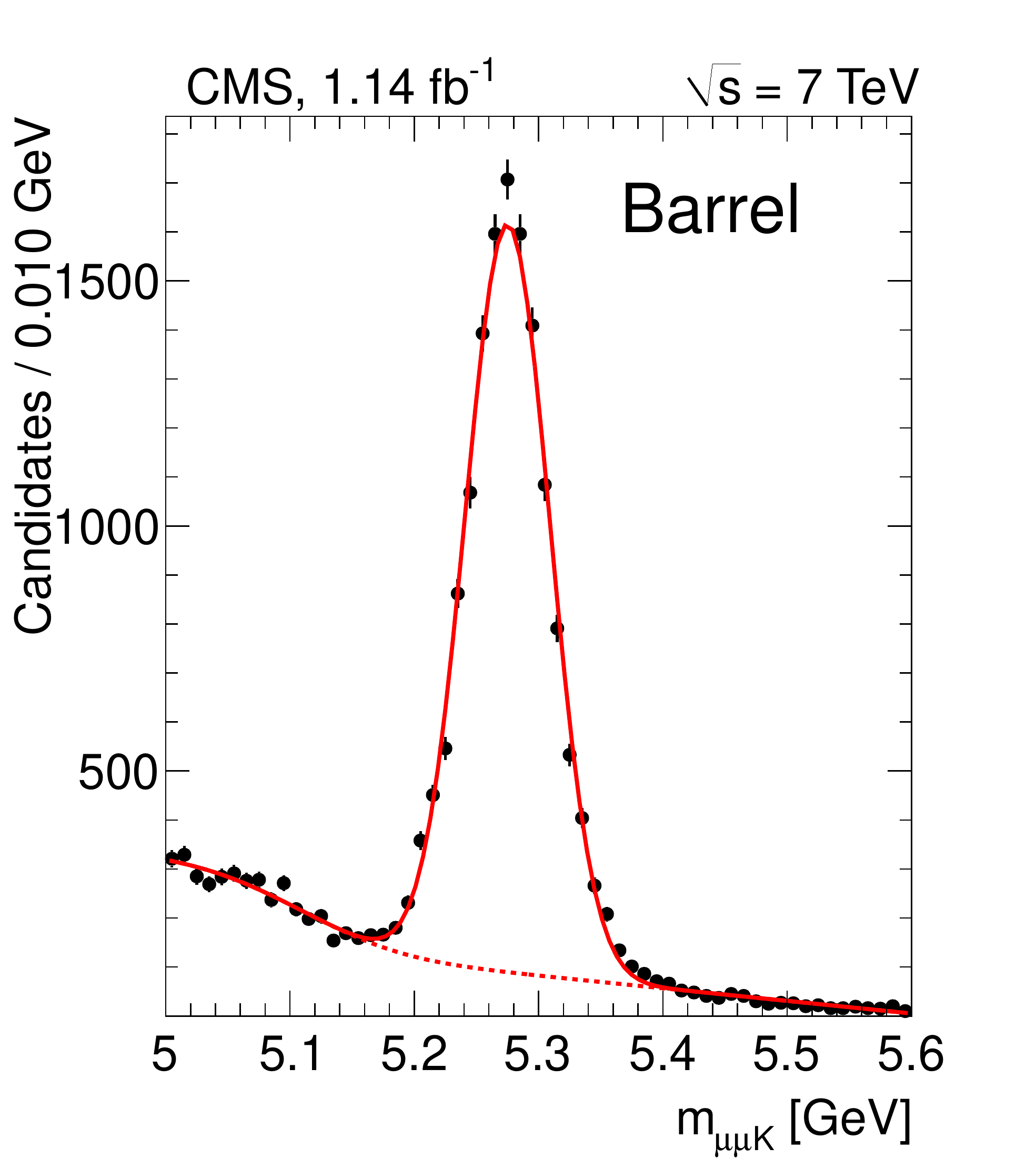}
\includegraphics[width=\figwid]{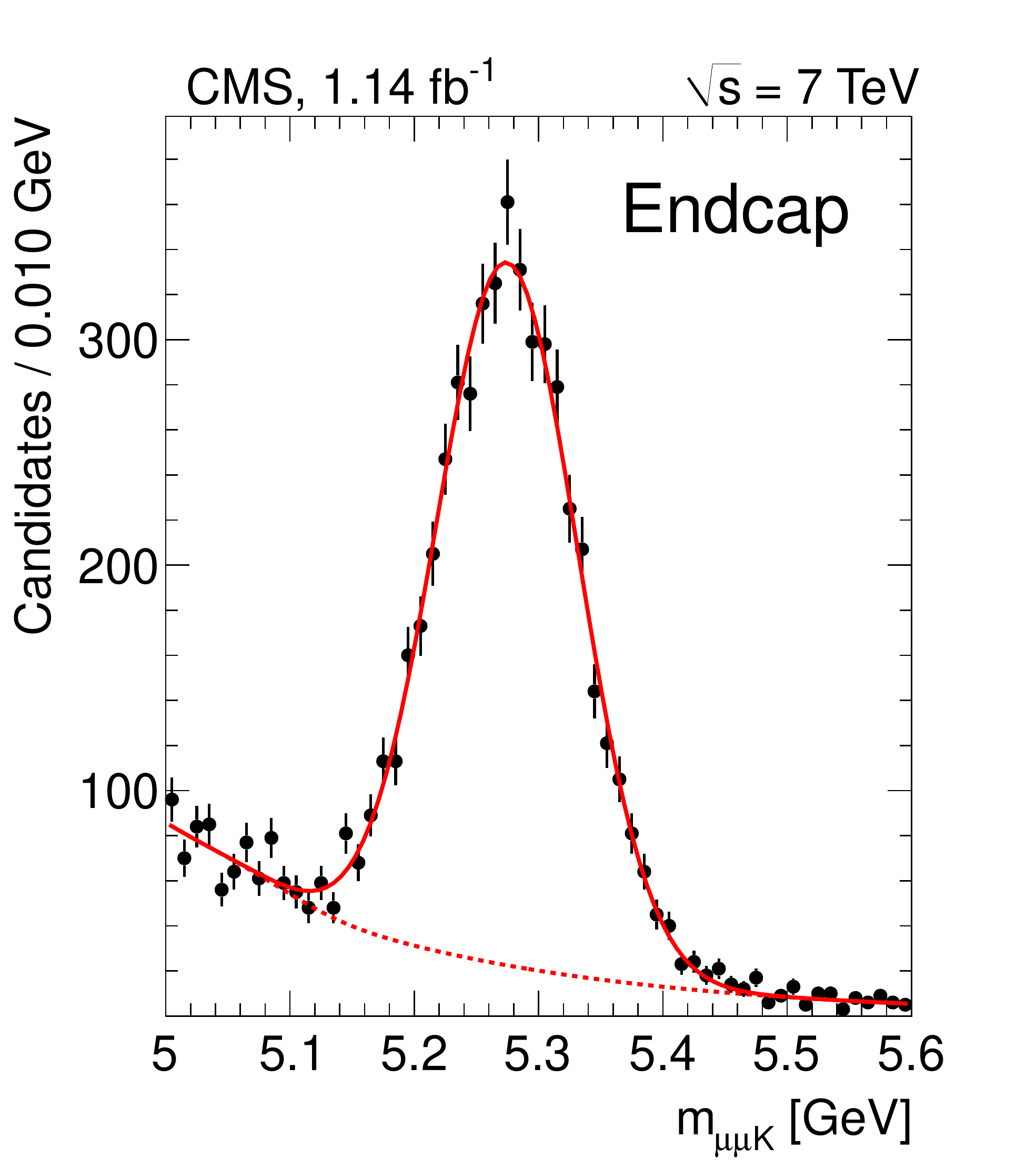}
   \caption{\bupsik\ invariant mass distributions in the barrel (left)
     and endcap (right) channels. The solid (dashed) lines show the fits
     to the data (background).}
   \label{fig:resNorm}
  \end{centering}
\end{figure}

The total efficiency for $\bupsik\to\mup\mun\Kp$, including the
detector acceptance, is $\varepsilon_{\mathrm{tot}}^{\Bp} = (7.7\pm0.8)\times10^{-4}$
and $(2.7\pm0.3)\times10^{-4}$, respectively for the
barrel and endcap channels,
where statistical and systematic uncertainties are combined.  The
acceptance has a systematic uncertainty of 4\%, estimated by comparing
the values obtained with different \bbbar\ production mechanisms
(gluon splitting, flavor excitation, and flavor creation).  The
uncertainty on the event selection efficiency for the \bupsik\
normalization sample is 4\%, evaluated from
differences between measured and simulated \bupsik\ events.  The
uncertainty on the signal efficiency (7.9\%) is evaluated using the
\bspsiphi\ control sample. The invariant mass distributions are fitted
with
a Gaussian function for the signal and an exponential
(barrel) or a first-degree polynomial (endcap) plus an error function
for the background, as shown in Fig.~\ref{fig:resNorm}. Applying the
same selection requirements as for the signal sample, the observed
number of \bupsik\ candidates in the barrel (endcap) channel is
$N_{\mathrm{obs}}^{\Bp} =
13\,045 \pm
\vuse{default-11:N-OBS-BPLUS0:sys} $
($\vuse{default-11:N-OBS-BPLUS1:val} \pm
\vuse{default-11:N-OBS-BPLUS1:sys} $).
The uncertainty includes a systematic term caused by
fit and background parametrization effects,
estimated to be 5\% from MC studies.

\renewcommand{\sample}{CsData-CsMc}
\renewcommand{\channel}{A}

To quantify a possible dependence on the pileup, the efficiencies of
the isolation and the flight length significance requirements are
calculated as functions of the number of reconstructed primary
vertices. No dependence is observed for
events with up to 12 primary vertices for the normalization and
control samples.

The \bsmm\ branching fraction is measured separately in the barrel
and endcap channels using
\begin{eqnarray}
  \cbf(\bsmm)
  &=&  \frac{N_\mathrm{S}}{N_{\mathrm{obs}}^{\Bp}} \,
  \frac{f_\cPqu}{f_\cPqs} \,
  \frac{\varepsilon_{\mathrm{tot}}^{\Bp}}{\varepsilon_{\mathrm{tot}}} \,
  \cbf(\Bp)\label{eq:schema},
\end{eqnarray}
and analogously for the \bdmm\ case, where $N_\mathrm{S}$ is the
background-subtracted number of observed \bsdmm\ candidates in the
signal window ($5.3 < m_{\mu\mu} < 5.45\gev$ for $\Bs$ and $5.2 <
m_{\mu\mu} < 5.3\GeV$ for \Bz) and $\varepsilon_\mathrm{tot}$ is the
total signal efficiency of all selection requirements.  The ratio of
the \Bs\ and \Bp\ meson production fractions  is $f_\cPqs/f_\cPqu = 0.282\pm0.037$ and $\cbf(\Bp) \equiv
\cbf(\Bp\to\jpsi\Kp\to\mup\mun\Kp) =
(6.0\pm0.2)\times10^{-5}$~\cite{Nakamura:2010zzi}. (We use $f_\cPqs =
0.113\pm 0.013$ and $f_\cPqu = 0.401\pm0.013$ from the main section of
Ref.~\cite{Nakamura:2010zzi} and account for the correlations in the
ratio.)

Events in the signal window can result from real signal decays,
combinatorial background, and ``peaking'' background from decays
of the type $\cPBds\to hh'$, where $h,h'$ are charged hadrons
misidentified as muons.  The expected number of signal events,
$N_{\mathrm{signal}}^{\mathrm{exp}}$, is calculated assuming the SM
branching fraction and is normalized to the \Bp\ yield.
The expected number of combinatorial background events, $N_{\mathrm{comb}}^{\mathrm{exp}} = 3 (4)$ in the barrel (endcap), is evaluated by interpolating to the signal window the number of events observed in the sideband regions.
The interpolation procedure assumes a flat
background shape and has a systematic uncertainty of 4\%, evaluated by
varying the flight length significance
selections and by using a floating slope.  The expected number of peaking
background events, $N_{\mathrm{peak}}^{\mathrm{exp}}$, is evaluated from
MC simulation and muon misidentification rates measured in
$\KS\to\pip\pim$, $\phi\to\Kp\Km$, and $\Lambda\to p\pim$
samples~\cite{misid:2010}.

\begin{figure}[htbp]
  \begin{centering}
\includegraphics[width=\figwid]{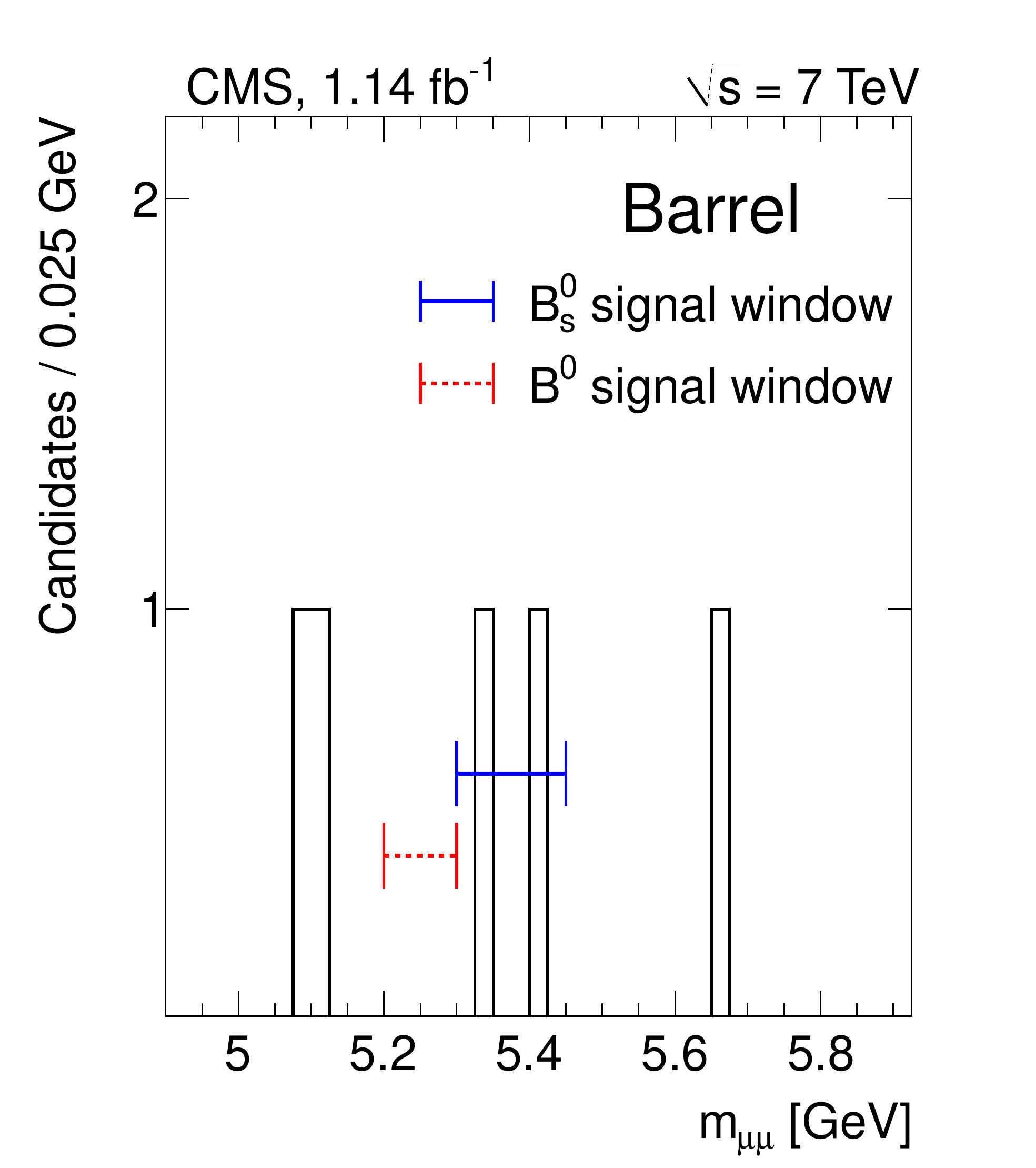}
\includegraphics[width=\figwid]{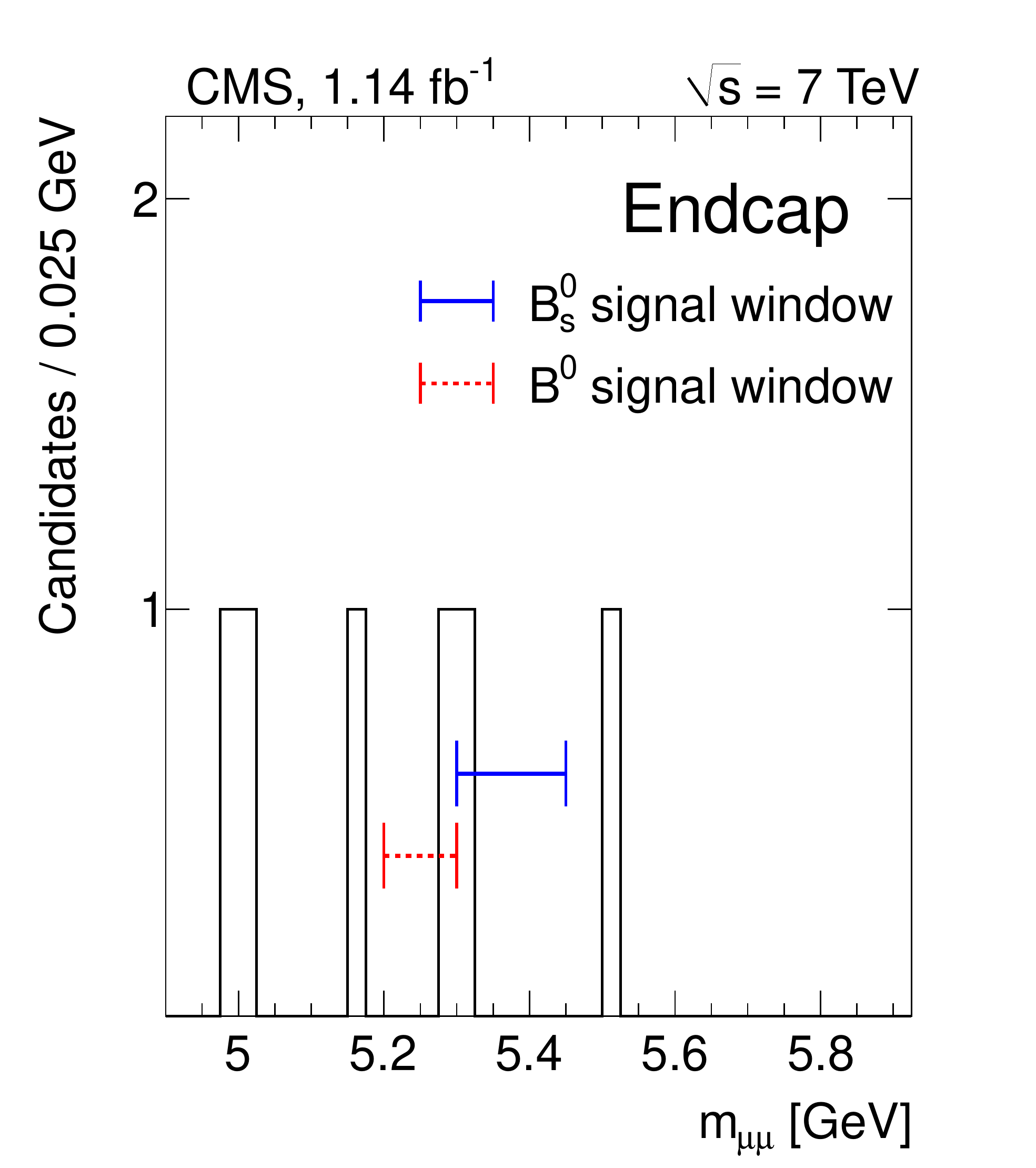}
   \caption{Dimuon invariant mass
     distributions in the barrel (left) and endcap (right)
     channels. The signal windows for \Bs\ and \Bz\
     are indicated by horizontal lines. }
   \label{fig:result}
  \end{centering}
\end{figure}

Figure~\ref{fig:result} shows the measured dimuon invariant mass
distributions. Three events are observed in the \bsmm\ signal windows
(two in the barrel and one in the endcap), while only one
event is observed in the \bdmm\ endcap channel. This observation is
consistent with the SM expectation for signal plus background.
Upper limits are determined with
the $\mathrm{CL_s}$ approach~\cite{Read:451614}.
Table~\ref{t:allTheNumbers} shows the values needed for the extraction of the results, separately for the barrel and endcap channels.  The obtained upper limits on the
branching fractions are ${\cal B}(\PBzs \to \mu^+ \mu^-) <
1.9\times10^{-8}$ $(1.6\times10^{-8})$ and ${\cal B}(\PBz \to \mu^+
\mu^-) < 4.6\times10^{-9}$ $(3.7\times10^{-9})$, at 95\% (90\%) CL.
The median expected upper limits at 95\% CL are $1.8\times10^{-8}$
($4.8\times10^{-9}$) for \bsmm (\bdmm). The background-only $p$ value
is 0.11 (0.40) for \bsmm\ (\bdmm), corresponding to $1.2$ ($0.27$)
standard deviations. The $p$ value is 0.053 when assuming a \bsmm\
signal at 5.6 times the SM value, as reported in
Ref.~\cite{CDF:2011fi}.

In summary, a search for the rare decays $\PBzs \to \mu^+ \mu^-$ and
$\PBz \to \mu^+ \mu^-$ has been performed on a data sample of $\Pp\Pp$
collisions at $\sqrt{\rm s}=7\,\mathrm{TeV}$ corresponding to an
integrated luminosity of $\vuse{lumi} \fbinv$.
The observed event
yields are consistent with those expected adding background and SM
signals. Upper limits on the branching fractions have been determined
at 90\% and 95\%\,CL.

\bigskip

We wish to congratulate our colleagues in the CERN accelerator
departments for the excellent performance of the LHC machine. We thank
the technical and administrative staff at CERN and other CMS
institutes, and acknowledge support from: FMSR (Austria); FNRS and FWO
(Belgium); CNPq, CAPES, FAPERJ, and FAPESP (Brazil); MES (Bulgaria);
CERN; CAS, MoST, and NSFC (China); COLCIENCIAS (Colombia); MSES
(Croatia); RPF (Cyprus); Academy of Sciences and NICPB (Estonia);
Academy of Finland, MEC, and HIP (Finland); CEA and CNRS/IN2P3
(France); BMBF, DFG, and HGF (Germany); GSRT (Greece); OTKA and NKTH
(Hungary); DAE and DST (India); IPM (Iran); SFI (Ireland); INFN
(Italy); NRF and WCU (Korea); LAS (Lithuania); CINVESTAV, CONACYT,
SEP, and UASLP-FAI (Mexico); MSI (New Zealand); PAEC (Pakistan); SCSR
(Poland); FCT (Portugal); JINR (Armenia, Belarus, Georgia, Ukraine,
Uzbekistan); MST, MAE and RFBR (Russia); MSTD (Serbia); MICINN and
CPAN (Spain); Swiss Funding Agencies (Switzerland); NSC (Taipei);
TUBITAK and TAEK (Turkey); STFC (United Kingdom); DOE and NSF (USA).

\clearpage

\bibliography{auto_generated}
\cleardoublepage \appendix\section{The CMS Collaboration \label{app:collab}}\begin{sloppypar}\hyphenpenalty=5000\widowpenalty=500\clubpenalty=5000\textbf{Yerevan Physics Institute,  Yerevan,  Armenia}\\*[0pt]
S.~Chatrchyan, V.~Khachatryan, A.M.~Sirunyan, A.~Tumasyan
\vskip\cmsinstskip
\textbf{Institut f\"{u}r Hochenergiephysik der OeAW,  Wien,  Austria}\\*[0pt]
W.~Adam, T.~Bergauer, M.~Dragicevic, J.~Er\"{o}, C.~Fabjan, M.~Friedl, R.~Fr\"{u}hwirth, V.M.~Ghete, J.~Hammer\cmsAuthorMark{1}, S.~H\"{a}nsel, M.~Hoch, N.~H\"{o}rmann, J.~Hrubec, M.~Jeitler, W.~Kiesenhofer, M.~Krammer, D.~Liko, I.~Mikulec, M.~Pernicka, B.~Rahbaran, H.~Rohringer, R.~Sch\"{o}fbeck, J.~Strauss, A.~Taurok, F.~Teischinger, C.~Trauner, P.~Wagner, W.~Waltenberger, G.~Walzel, E.~Widl, C.-E.~Wulz
% ...
\vskip\cmsinstskip
% ...
1:~~Also at CERN, European Organization for Nuclear Research, Geneva, Switzerland\\
% ...

\end{sloppypar}
\end{document}